\pdfoutput=1

\documentclass[apj]{emulateapj}
\usepackage{apjfonts}
\usepackage[bookmarks=true,colorlinks=true,citecolor=blue,linkcolor=magenta,urlcolor=cyan]{hyperref}

\usepackage{natbib}
\usepackage{amsmath}
\usepackage{amssymb}
\usepackage{subfigure}
\usepackage{url}
\usepackage{CJK}

\renewcommand{\vec}[1]{ {\mathbf #1} }

\newcommand{\Fig}{{Figure}}

\slugcomment{Accepted for publication in ApJL}
\shorttitle{Simulation of a Sigmoid Eruption}
\shortauthors{Jiang et al.}

\begin{document}
\begin{CJK*}{UTF8}{gbsn}

\title{MHD Simulation of a Sigmoid Eruption of Active Region 11283}

\author{
  Chaowei Jiang (江朝伟)\altaffilmark{1},
  Xueshang Feng (冯学尚)\altaffilmark{1},
  S.~T. Wu (吴式灿)\altaffilmark{2},
  Qiang Hu (胡强)\altaffilmark{2}}
\email{cwjiang@spaceweather.ac.cn, fengx@spaceweather.ac.cn}

\altaffiltext{1}{SIGMA Weather Group, State Key Laboratory for Space
  Weather, Center for Space Science and Applied Research, Chinese
  Academy of Sciences, Beijing 100190}

\email{wus@uah.edu, qh0001@uah.edu}

\altaffiltext{2}{Center for Space Plasma and Aeronomic Research, The
  University of Alabama in Huntsville, Huntsville, AL 35899, USA}

\begin{abstract}
  Current magnetohydrodynamic (MHD) simulations of the initiation of
  solar eruptions are still commonly carried out with idealized
  magnetic field models, whereas the realistic coronal field prior to
  eruptions can possibly be reconstructed from the observable
  photospheric field. Using a nonlinear force-free field extrapolation
  prior to a sigmoid eruption in AR 11283 as the initial condition in
  a MHD model, we successfully simulate the realistic initiation
  process of the eruption event, as is confirmed by a remarkable
  resemblance to the SDO/AIA observations. Analysis of the
    pre-eruption field reveals that the envelope flux of the sigmoidal
    core contains a coronal null and furthermore the flux rope is
    prone to a torus instability. Observations suggest that
    reconnection at the null cuts overlying tethers and likely
    triggers the torus instability of the flux rope, which results in
    the eruption.
  This kind of simulation demonstrates the capability of modeling the
  realistic solar eruptions to provide the initiation process.
\end{abstract}


\section{Introduction}
\label{sec:intro}

Solar eruptions are major drivers of the space weather, and
the key of forecast of the space weather is to understand the eruption
mechanism. Although manifested as different observational forms
including the flares, filaments eruptions and coronal mass ejections
(CMEs), it is commonly accepted that solar eruptions are caused by
the disruption of the coronal magnetic field, in which the magnetic
free energy stored in the corona prior to the event is
released. However, the mechanism of their initiations is still
unclear. A variety of theoretical models have been proposed to explain
the initiation of solar eruptions \citep[see, e.g.,][and references
therein]{Forbes2006, Schmieder2013}.

Numerical simulations are powerful tool to constrain the theoretical
models. Over the past few years, many authors have carried out
different MHD simulations to investigate the initiation process of
eruptions \citep[e.g.,][etc.]{Amari2003, MacNeice2004, Torok2005,
  Aulanier2009, Fan2010, Kliem2010, Roussev2012}, which have greatly
improved our knowledge of the eruption mechanism.  However, these
simulations commonly involve highly idealized magnetic-field
configuration with perfect symmetry and smoothness (a well known
example is \citet{Titov1999}'s flux rope model, which is used by
\citet{Torok2005, Torok2011, Torok2013}) which only mimics the coronal
field. Taking into account a more realistic coronal environment might
affect the modeling result significantly.

Very recently modelers have developed non-symmetric versions of the
idealized models to improve their abilities of resembling the real
observations \citep{Aulanier2010, Aulanier2012, FanY2011, Torok2011,
  Torok2013, Zuccarello2012}, but even this is only qualitative. The
realistic coronal field might be much more complex as implied by
complex flux distribution of the observed photospheric field, and is
difficult to recover by the idealized models. A step forward of
understanding what really happens in the solar eruptions, certainly
necessitates the numerical simulations constrained directly by the
observations (if available). Only in this way, the eruption process
can then be expected to be reproduced accurately, and critical
parameters, e.g., the eruption direction and speed, can be hoped to be
achieved correctly. Although there are also MHD simulation works using
line-of-sight observed magnetograms as boundary conditions
\citep[e.g.,][]{Lugaz2011, Downs2012}, these simulations focus on the
large-scale field response to the eruptions and the propagation of the
CMEs while their initiation mechanisms are not treated properly.

At the moment, the photospheric vector mangetograms combined with the
force-free field model have been used successfully to reconstruct the
realistic coronal field prior to eruptions, when the coronal field is
near in equilibrium and thus can be assumed to be force-free
\citep{wiegelmann2012solar}. Critical structures, e.g., twisted flux
rope (FR) and 3D coronal null point, which are basic building block of
many eruption models \citep[e.g., ][]{Forbes1991, Kliem2006PRL,
  Antiochos1999}, can be reconstructed by using the general nonlinear
force-free field (NLFFF) model \citep{Guo2010, Canou2010,
  Sun2012b, Jiang2013NLFFF}. Moreover, based on the NLFFF
extrapolation for the pre-eruption field, magnetic topology study can
shed important light on the eruption mechanism \citep{Savcheva2012a,
  Sun2012b}. This inspires us to input the NLFFF based on real
magnetograms into MHD simulation for a better modeling of the real
eruption other than using idealized models.

In this Letter, we report a successful simulation of the initiation
process of a sigmoid eruption in AR~11283 in its realistic magnetic
configuration. The MHD model is initialized by a NLFFF extrapolation
for the pre-eruption field,
and reproduces almost accurately the SDO/AIA observations of the
eruption initiation process.

\begin{figure*}[htbp]
  \centering
  \includegraphics[width=\textwidth]{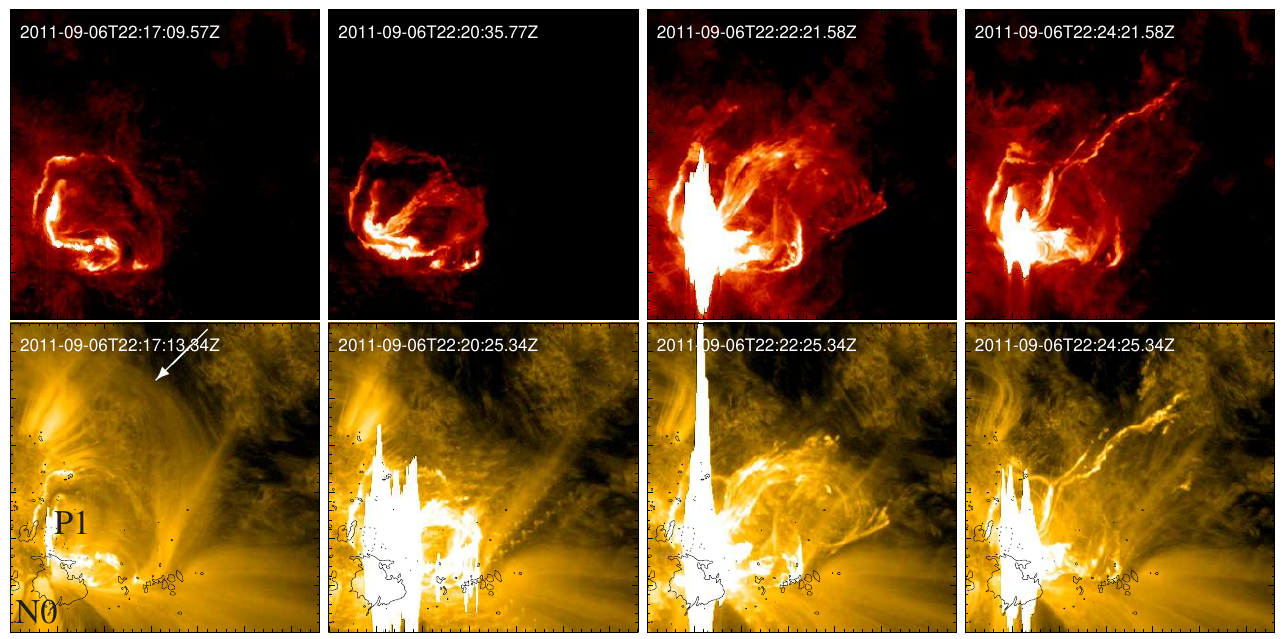}
  \caption{SDO/AIA observations (in 304 and 171 {\AA}) of the sigmoid
    eruption in AR~11283 on 2011 September 6. The contours overlying
    the images are the line-of-sight magnetic field with $\pm
    500$~G. {\it An animation of this figure is available.}}
  \label{fig:1}
\end{figure*}

\begin{figure}[htbp]
  \centering
  \includegraphics[width=0.45\textwidth]{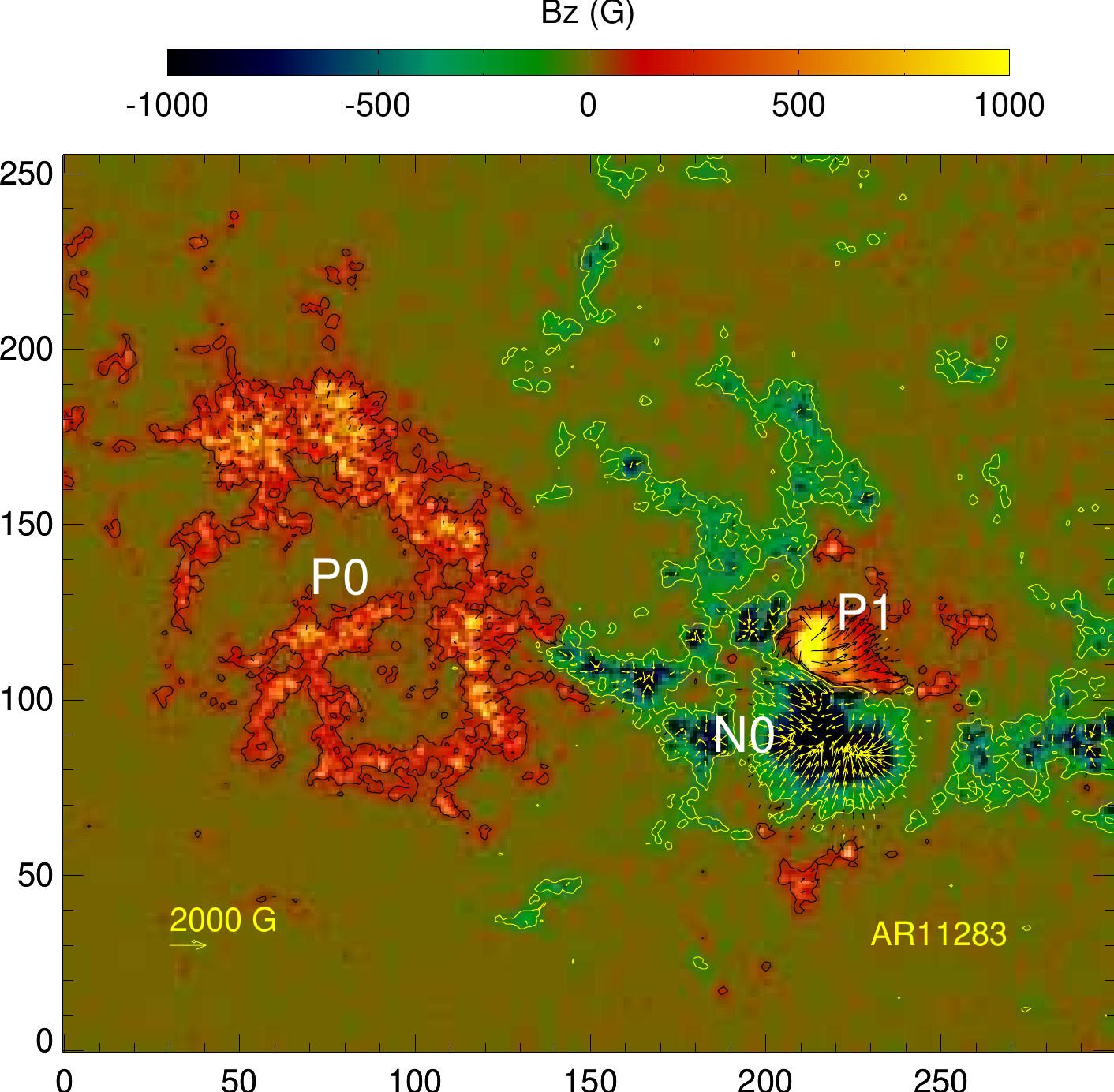}
  \caption{The vector magnetogram of AR 11283 taken by SDO/HMI at
    22:00~UT, 2011 September 6. The length unit is one arcsec, the
    contour lines represents $B_z$ of $\pm 100$~G and the vectors
    represent the transverse field (above 200~G). There are three main
    polarities labeled as P0, N0 and P1.}
  \label{fig:2}
\end{figure}

\section{Observations}
\label{sec:obse}

Our target AR 11283 is a very eruptive source region, producing
several flares and CMEs when it was located near the disk center. We
focus here on the initiation process of a CME on 2011 September 6,
i.e., a sigmoid eruption around 22:20~UT (see \Fig~1). At 22:00~UT the
coronal magnetic field was still near-static and a sigmoid was clearly
observed in AIA-94 channel (see \Fig~3). After then the field evolved
slightly until its drastic eruption at 22:18~UT, which is observed as
a fast rising of a FR with a duration of 6~min. The evolution of the
FR is most clearly seen in AIA 304 and 171 channels (shown in the
first two rows of \Fig~1). The FR rose with a speed of roughly
300~km/s in a non-radial direction approximately toward the northwest
(of the disk plane). It developed a highly asymmetric and helical
shape, with the north leg much brighter than the south leg. Inspection
of the successive eruption progress shows that the north leg (rooted
in the positive polarity, labeled as P1 in \Fig~2) of the FR was held
fixed while the south footpoint (rooted in N0) appears to slip along
the west during the rising of the rope. Along with the eruption was a
X2.1 flare (started at 22:12~UT and peaked at 22:20~UT), with a
remarkably circular ribbon, which strongly indicates the presence of a
magnetic null configuration \citep{Masson2009,Wang2012}, surrounding
the footpoint site of the eruption (see \Fig~3). The circular flare
ribbon, as a signature of reconnection at the null, is observed a
little earlier than the rising of the FR. We also note that a
  group of large closed loops above the eruption site (denoted by the
  arrow on \Fig~1) expanded and opened quickly with the onset of the
  flare before the rising of the FR.

\Fig~2 shows a vector magnetogram\footnote{The magnetogram is
  downloaded from website
  \url{http://jsoc.stanford.edu/jsocwiki/ReleaseNotes2}, where
  products of HMI vector magnetic field datas are released for several
  ARs.} of AR 11283 at 22:00~UT taken by HMI \citep{Schou2012}. Three
main polarities are labeled as P0, N0, and P1. Clearly P0 and N0 are
much more dispersed than P1, since they were preexisting longer than
P1 which emerged into N0 only after September 4. The flare and
eruption took place near the polarity inversion line (PIL) between N0
and P1, where non-potential energy is stored by the strongly sheared
field. As a parasitic polarity of N0, P1 is surrounded by the negative
flux. Such magnetic flux distribution also suggests the existence of a
null.
Our following numerical models use
this magnetogram as input. Unfortunately, this magnetogram is limited
to a too small field of view (FoV) to model the global magnetic
environment for the eruption, and even worse the FR erupts towards the
west and out of magnetogram's FoV after 22:22~UT. As a result, we can
only simulate the very early phase of the eruption process, e.g., how
the AR's local confinement is broken, and still fortunately, this
early phase is captured by AIA, with which the simulation can be
compared. 

\begin{figure}[htbp]
  \centering
  \includegraphics[width=0.45\textwidth]{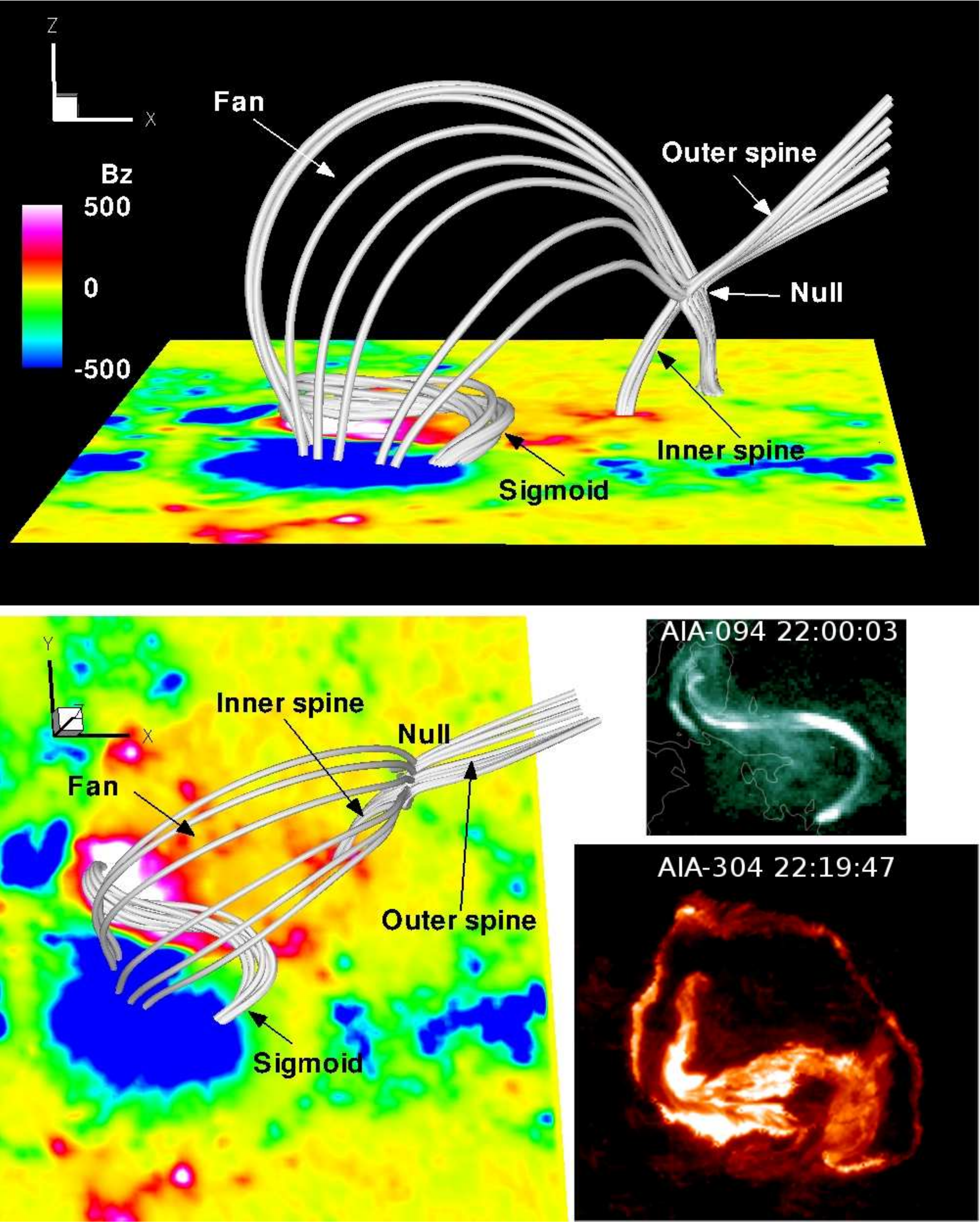}
  \caption{Magnetic topology based on NLFFF extrapolation for the
    pre-eruption field. The sigmoid field is the low-lying S-shaped
    lines; field lines closely touching the null outline the spine-fan
    topology of the null, where the lines form a X-point
    configuration. The null point locates about 18~arcsec (13~Mm) above the
    photosphere and about 50~arcsec away from the sigmoid in the same
    direction of the eruption. The upper panel is a side view and the
    bottom is the SDO view. Compared on the bottom right are the
    sigmoid observed in SDO/AIA-94 channel and the circular flare
    ribbon in AIA-304. }
  \label{fig:3}
\end{figure}

\begin{figure*}[htbp]
  \centering
  \includegraphics[width=\textwidth]{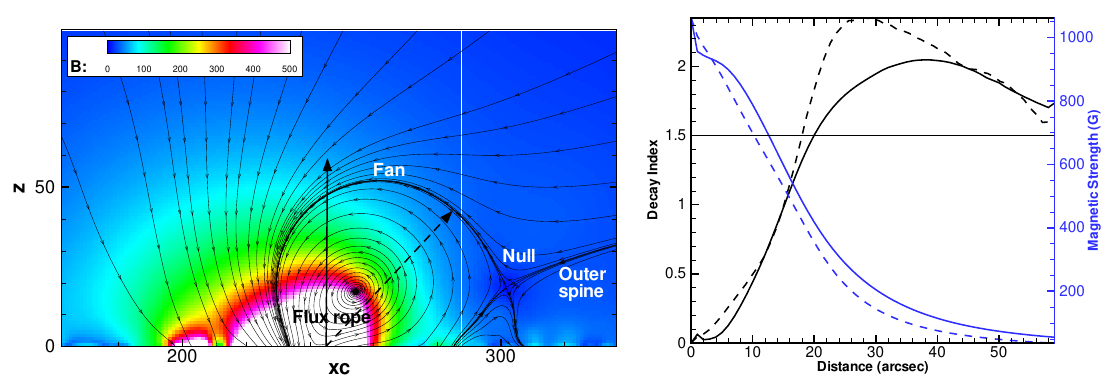}
  \caption{A vertical cross section cutting through the coronal
    null and the middle of the sigmoid, thus roughly parallel to the
    direction of eruption as observed. Streamlines are the projection
    of the 3D field lines and the field strength $B$ is imaged on the
    background. In the core region above the PIL, a spiral forms with
    center at the axis of a FR (with height of about
    16~arcsec).
    The thick solid and dashed lines with arrows denotes two
    paths (started from near the PIL at the bottom) along which the
    the field strength and its decay indexes are plotted in the right
    panel. The solid one is radial and the dashed roughly the eruption
    direction modeled by the MHD simulation.
    Note that the FR reaches almost 20~arcsec, a height with decay
    index $>1.5$.}
 \label{fig:slice}
\end{figure*}


\section{The Pre-Eruption Magnetic Field}
\label{sec:nlfff}

Considering that the coronal magnetic field in static is well modeled
by the nonlinear force-free field (NLFFF) model, we have used our
CESE--MHD--NLFFF
code \citep{Jiang2012apj, Jiang2012apj1} to extrapolate the
pre-eruption 3D field at 22:00~UT from the photospheric vector
magnetogram as shown in \Fig~2. A detailed implementation of this
extrapolation has been described by \citet{Jiang2013NLFFF}, where we
show that the large-scale coronal loops are recovered well by the
NLFFF solution, and especially the sigmoid is resembled very closely
by a bundle of S-shaped field lines near the main PIL of N0-P1. Here
we briefly present the basic magnetic topology and review some special
aspects of the NLFFF solution, which is important for our
interpretation and MHD simulation of the eruption.

\Fig~3 and a cross section in \Fig~\ref{fig:slice} show the basic
structure of the NLFFF. In \Fig~3, the low-lying S-shaped field lines
resemble precisely the AIA-94 observation of the sigmoid. Closely
above the sigmoid is a FR (see \Fig~\ref{fig:slice}), which is
strongly sheared but slightly twisted (excluding the kink
instability). The core field carries a strong field-aligned current
and stores a free energy content of nearly $10^{32}$~erg, which is
sufficient for a major eruption. Overlying the stressed core is the
closed flux playing the role of tethers that constrain the
pre-eruptive sigmoidal field. Indeed there is a 3D magnetic null point
associated with the overlying field. The null locates in the northwest
(the direction of the eruption) of the sigmoid with a height of
13~Mm. In the figure we plot several representative field lines
closely touching the null, which forms a remarkable X-point
configuration at the null (see also \Fig~\ref{fig:slice}). These field
lines outline the so-called spine-fan topology of a coronal null
\citep{Lau1990, Torok2009}, as labeled in the figure.
Naturally, the fan surface intersects with the
chromosphere as a closed circle. The circular flare ribbon is an
evidence of reconnection occurring at the null, and is produced by
reconnection-accelerated particles chasing along the fan lines down to
the chromosphere.

Such field configuration may be unstable as the null reconnection
cuts the overlying tethers of the sigmoidal core
and facilitates the outward expansion of the FR.
Similar configurations, i.e., low-lying sheared/twisted core confined
by overlying flux associating with a null
were frequently found,
e.g., \citet{Lugaz2011, Sun2012b}, and \Fig~7 of the latter can also
be used here to illustrate how the null reconnection works to open the
overlying flux. However, the null reconnection does not necessarily
leads to a successful eruption, if the overlying envelop flux, even
partially cut by reconnection, is still strong enough to confine the
stressed core. We thus further study the decay speed of the envelop
flux, quantified by a decay index ($=-(r/B)(\partial B/\partial r)$),
to see whether a torus instability (TI) exists here for the
FR/overlying flux system \citep{Kliem2006PRL, Torok2007}. In
\Fig~\ref{fig:slice}, the decay indexes along two different directions
from near the PIL at the bottom (shown in the left panel) are plotted
(in the right panel). Clearly in both directions, the FR reaches {\it
  almost (but not fully)} the domain of TI, in which the decay index
is above the critical threshold of 1.5 for TI as found by
\citep[e.g,][]{Torok2007, Aulanier2009}. Furthermore, the field decays
faster in the non-radial direction than in the radial direction (also
can be seen in the field strength image in \Fig~\ref{fig:slice}), thus
can result effectively a non-radial eruption.

Hence the eruption appears to be caused jointly by tether cutting at
the null and TI. Specifically, the null reconnection first cuts some
overlying tethers, resulting in a small expansion of the FR, which
thus enters into the unstable regime of TI; the TI then drives the
explosion of the FR, producing a fast evolution due to positive
feedback between the tether-cutting reconnection and the expansion of
the FR driven by TI. In the next section we put the extrapolated NLFFF
into a MHD model along with a plasma in hydrostatic equilibrium to
test such eruption mechanism. Before this, an important remark should
be made that the NLFFF solution is not exactly force-free but with
small numerical errors,
thus there are small residual Lorentz forces \citep{Jiang2013NLFFF}.
The residual force can play the role of perturbation to the unstable
system by inducing small velocity and triggering reconnection at the
null.
Also we note that if the NLFFF system is stable, it will just relax to
a magnetohydrostatic equilibrium in the MHD computation with balance
between Lorentz force and plasma.

\begin{figure*}[htbp]
  \centering
  \includegraphics[width=\textwidth]{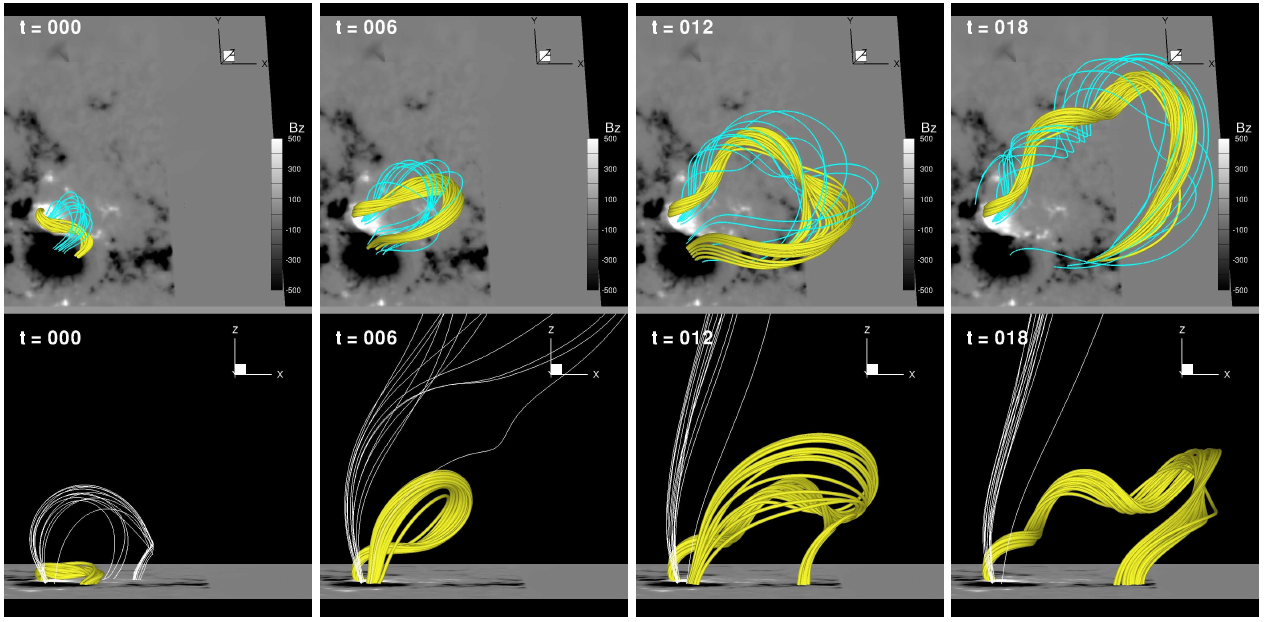}
  \caption{MHD simulation of the eruption. Viewing angle in the upper
    row is aligned with the SDO/AIA observations, while the bottom row
    is side view from south. Two sets of the core field lines (cyan
    and yellow) are shown, which are traced from the same positive
    polarity at the bottom. Another set of lines that initially touch
    the null point are traced from the negative polarity to show their
    reconnection and opening. The bottom magnetogram of $B_z$ is also
    shown. Note that $B_{z}$ is set to zero near the west boundary,
    where the computational volume is out of the FoV of the vector
    magnetogram.}
 \label{fig:MHD}
\end{figure*}

\section{MHD Modeling of the Eruption}
\label{sec:mhd}

The MHD simulation is carried out using our recently developed
CESE--MHD model \citep{Jiang2012c}. Here we solve a full set of 3D
time-dependent MHD equations with consideration of solar gravity. No
explicit resistivity is included in the magnetic induction equation,
and magnetic reconnection is still allowed due to numerical diffusion
if current sheets are thin enough and under the grid resolution
\citep[e.g.,][]{Torok2005}. A small kinematic viscosity is used, which
may reduce the evolution speed of eruption, but is usually necessary
for numerical stability in the computation of an extremely dynamical
process. The modeling volume is a rectangle box of 224~arcsec $\times$
256~arcsec $\times$ 256~arcsec, (which is a sub-volume of a larger
NLFFF extrapolation box) with our region of interest at the center of
the box. A non-uniform mesh is used with the smallest grid
$0.5$~arcsec near the bottom (the photosphere) and the largest grid
4~arcsec along the side and top faces of the box. The initial
conditions of the simulation consists of the NLFFF solution described
in Section~3 and a simplified plasma similar to \citep{Jiang2012c}:
the plasma is uniform horizontally and in hydrostatic equilibrium
vertically with a constant temperature in the gravitational field. The
plasma is configured with low $\beta$ (i.e., the ratio of plasma
pressure to magnetic pressure, the samllest value is $2\times
10^{-4}$) and high Alfv{\'e}n speed (highest value of $10^{4}$~km/s)
to model the coronal environment \citep[see the profiles of the
parameters in \Fig~2 of][]{Jiang2012c}.  The non-reflecting
projected-characteristic method \citep[see][and reference
therein]{Jiang2011, Wu2006} is implemented at the side and top
boundaries of the computational box. At the bottom surface, all the
variables are simply fixed to model the line-tied effect of the
high-$\beta$ and dense photosphere. This is reasonable since the
eruption is very fast with a short duration of less than 10 minutes
during which the photospheric driving effect (e.g., the surface motion
and flux emergence) can be neglected. Comparing vector magnetgrams
taken at 22:00~UT and 22:24~UT shows that the change is indeed small.
The computation is stopped when the expansion of the core field
reaches the right side boundary, a bit far away from the edge of FoV
of the bottom magnetogram. The end time is at $t=18\tau$ where the
time unit $\tau$ is scaled as 20~s.

The MHD results shown in \Fig~5 are compared with the AIA observations
in \Fig~1. To show the basic process of the eruption, we plot two sets
of field lines, the core field lines and the overlying field lines
that initially touch the null point. The core field lines are traced
starting from polarity P1, since observation suggests the footpoint of
the FR is fixed there. As expected, the overlying flux reconnects and
opens and the core field expands and rises rapidly, which confirms the
eruption mechanism interpreted in Section~3. The perfect resemblance
of the simulation with observations can be seen.  Both the development
of helical shape and the rising direction of the FR are reproduced
almost precisely at each time snapshot as shown. The simulation
provides a full 3D picture of the evolution.

The eruption is modulated strongly by the highly-anisotropic ambient
field. It is due to the location of the reconnection at the null that
leads a strongly inclined non-radial path of the eruption as seen both
from the SDO and side views. The observed asymmetry, e.g., that the
north leg of the FR is much brighter than the south leg, can be
explained well by the simulated field: twist/writhe of the north-leg
field is remarkably stronger than the south leg, and thus during the
eruption the north leg can hold more plasma by dips due to the twist;
moreover the south leg is more vertical (see the side view of the MHD
results in \Fig~1) and the plasma can easily flow back down by
gravity.

Even though magnetic field is fixed at the photosphere, the footpoints
of the erupting flux rope are not necessarily fixed there,
but instead, they can slip in the same polarity, as is suggested
by the observations (see Section~2). The MHD result shows that the
south footpoint has moved to the west edge of polarity N0 (in the FoV
of the magnetogram) at the end of the computation, which repeats the
observed slipping movement. The footpoint slipping can be attributed
to the so-called slip-running reconnection in the quasi-separatrix
layers \citep{Aulanier2006} between the FR field and its overlying
flux, where very thin current sheet can develop by the stress of the
flux expansion.

\section{Conclusions}
\label{sec:colu}

We have carried out a MHD simulation of a sigmoid eruption in a
realistic coronal magnetic environment without artificially tuned
parameter. Using a CESE--MHD--NLFFF code we first extrapolate the
static coronal field prior to eruption. Study of the pre-eruption
field shows that there is a coronal null related with the envelope
field overlying the sigmoidal core,
and the initiation of eruption can be explained by tether cutting at
the null triggering a TI of the FR--overlying field system.
The NLFFF is then input into a time-dependent MHD model to generate
the eruption process. Direct comparison of the eruption field with the
SDO/AIA images shows an excellent resemblance, demonstrating our
ability of modeling the realistic eruptions.
Due to the limited FoV of the magnetogram and model box based on
Cartesian geometry, the present simulation can only model how the
sigmoidal core breaks its local AR's confinement, while the explosion
out of the large and global field requires including of larger
computational volume and even the global corona-solar wind under
spherical geometry \citep{Lugaz2011, Feng2012apj}.

To our best knowledge, this is probably the first full-MHD simulation
of realistic initiation process of solar eruptions. With the success
of this simulation, more endeavors of the same kind, i.e., numerical
modeling the eruption events in a {\it quantitative} way and
comparable with observations {\it directly}, are inspired to resolve
the long-standing problem how solar eruptions are triggered and
driven.


\acknowledgments

We thank the anonymous referee for helpful comments on the
manuscript. This work is jointly supported by the 973 program under
grant 2012CB825601, the Chinese Academy of Sciences (KZZD-EW-01-4),
the National Natural Science Foundation of China (41204126, 41231068,
41274192, 41031066, and 41074122), and the Specialized Research Fund
for State Key Laboratories. The work performed by STW is supported by
NSF-AGS1153323. QH acknowledges NSF-AGS1062050 for partial support. 
The numerical calculation has been completed on our
SIGMA Cluster computing system. Data from observations are courtesy of
NASA/{SDO} and the HMI science teams.


\begin{thebibliography}{42}
\expandafter\ifx\csname natexlab\endcsname\relax\def\natexlab#1{#1}\fi

\bibitem[{{Amari} {et~al.}(2003){Amari}, {Luciani}, {Aly}, {Mikic}, \&
  {Linker}}]{Amari2003}
{Amari}, T., {Luciani}, J.~F., {Aly}, J.~J., {Mikic}, Z., \& {Linker}, J. 2003,
  \apj, 585, 1073

\bibitem[{{Antiochos} {et~al.}(1999){Antiochos}, {DeVore}, \&
  {Klimchuk}}]{Antiochos1999}
{Antiochos}, S.~K., {DeVore}, C.~R., \& {Klimchuk}, J.~A. 1999, \apj, 510, 485

\bibitem[{{Aulanier} {et~al.}(2012){Aulanier}, {Janvier}, \&
  {Schmieder}}]{Aulanier2012}
{Aulanier}, G., {Janvier}, M., \& {Schmieder}, B. 2012, \aap, 543, A110

\bibitem[{{Aulanier} {et~al.}(2006){Aulanier}, {Pariat}, {D{\'e}moulin}, \&
  {DeVore}}]{Aulanier2006}
{Aulanier}, G., {Pariat}, E., {D{\'e}moulin}, P., \& {DeVore}, C.~R. 2006,
  \solphys, 238, 347

\bibitem[{Aulanier {et~al.}(2009)Aulanier, T{\"o}r{\"o}k, D{\'e}moulin, \&
  DeLuca}]{Aulanier2009}
Aulanier, G., T{\"o}r{\"o}k, T., D{\'e}moulin, P., \& DeLuca, E.~E. 2009, \apj,
  708, 314

\bibitem[{{Aulanier} {et~al.}(2010){Aulanier}, {T{\"o}r{\"o}k}, {D{\'e}moulin},
  \& {DeLuca}}]{Aulanier2010}
{Aulanier}, G., {T{\"o}r{\"o}k}, T., {D{\'e}moulin}, P., \& {DeLuca}, E.~E.
  2010, \apj, 708, 314

\bibitem[{{Canou} \& {Amari}(2010)}]{Canou2010}
{Canou}, A. \& {Amari}, T. 2010, \apj, 715, 1566

\bibitem[{{Downs} {et~al.}(2012){Downs}, {Roussev}, {van der Holst}, {Lugaz},
  \& {Sokolov}}]{Downs2012}
{Downs}, C., {Roussev}, I.~I., {van der Holst}, B., {Lugaz}, N., \& {Sokolov},
  I.~V. 2012, \apj, 750, 134

\bibitem[{Fan(2010)}]{Fan2010}
Fan, Y. 2010, \apj, 719, 728

\bibitem[{{Fan}(2011)}]{FanY2011}
{Fan}, Y. 2011, \apj, 740, 68

\bibitem[{Feng {et~al.}(2012)Feng, Jiang, Xiang, Zhao, \& Wu}]{Feng2012apj}
Feng, X., Jiang, C., Xiang, C., Zhao, X., \& Wu, S. 2012, The Astrophysical
  Journal, 758, 62

\bibitem[{{Forbes} \& {Isenberg}(1991)}]{Forbes1991}
{Forbes}, T.~G. \& {Isenberg}, P.~A. 1991, \apj, 373, 294

\bibitem[{{Forbes} {et~al.}(2006){Forbes}, {Linker}, {Chen}, {Cid}, {K{\'o}ta},
  {Lee}, {Mann}, {Miki{\'c}}, {Potgieter}, {Schmidt}, {Siscoe}, {Vainio},
  {Antiochos}, \& {Riley}}]{Forbes2006}
{Forbes}, T.~G., {Linker}, J.~A., {Chen}, J., {Cid}, C., {K{\'o}ta}, J., {Lee},
  M.~A., {Mann}, G., {Miki{\'c}}, Z., {Potgieter}, M.~S., {Schmidt}, J.~M.,
  {Siscoe}, G.~L., {Vainio}, R., {Antiochos}, S.~K., \& {Riley}, P. 2006, \ssr,
  123, 251

\bibitem[{{Guo} {et~al.}(2010){Guo}, {Schmieder}, {D{\'e}moulin}, {Wiegelmann},
  {Aulanier}, {T{\"o}r{\"o}k}, \& {Bommier}}]{Guo2010}
{Guo}, Y., {Schmieder}, B., {D{\'e}moulin}, P., {Wiegelmann}, T., {Aulanier},
  G., {T{\"o}r{\"o}k}, T., \& {Bommier}, V. 2010, \apj, 714, 343

\bibitem[{{Jiang} {et~al.}(2011){Jiang}, {Feng}, {Fan}, \& {Xiang}}]{Jiang2011}
{Jiang}, C., {Feng}, X., {Fan}, Y., \& {Xiang}, C. 2011, \apj, 727, 101

\bibitem[{{Jiang} {et~al.}(2012{\natexlab{a}}){Jiang}, {Feng}, {Wu}, \&
  {Hu}}]{Jiang2012c}
{Jiang}, C., {Feng}, X., {Wu}, S.~T., \& {Hu}, Q. 2012{\natexlab{a}}, \apj,
  759, 85

\bibitem[{{Jiang} {et~al.}(2012{\natexlab{b}}){Jiang}, {Feng}, \&
  {Xiang}}]{Jiang2012apj1}
{Jiang}, C., {Feng}, X., \& {Xiang}, C. 2012{\natexlab{b}}, \apj, 755, 62

\bibitem[{{Jiang} \& {Feng}(2012)}]{Jiang2012apj}
{Jiang}, C. \& {Feng}, X. 2012, \apj, 749, 135

\bibitem[{{Jiang} \& {Feng}(2013)}]{Jiang2013NLFFF}
---. 2013, \apj, 769, 144

\bibitem[{{Kliem} {et~al.}(2010){Kliem}, {Linton}, {T{\"o}r{\"o}k}, \&
  {Karlick{\'y}}}]{Kliem2010}
{Kliem}, B., {Linton}, M.~G., {T{\"o}r{\"o}k}, T., \& {Karlick{\'y}}, M. 2010,
  \solphys, 266, 91

\bibitem[{{Kliem} \& {T{\"o}r{\"o}k}(2006)}]{Kliem2006PRL}
{Kliem}, B. \& {T{\"o}r{\"o}k}, T. 2006, Physical Review Letters, 96, 255002

\bibitem[{{Lau} \& {Finn}(1990)}]{Lau1990}
{Lau}, Y.-T. \& {Finn}, J.~M. 1990, \apj, 350, 672

\bibitem[{{Lugaz} {et~al.}(2011){Lugaz}, {Downs}, {Shibata}, {Roussev}, {Asai},
  \& {Gombosi}}]{Lugaz2011}
{Lugaz}, N., {Downs}, C., {Shibata}, K., {Roussev}, I.~I., {Asai}, A., \&
  {Gombosi}, T.~I. 2011, \apj, 738, 127

\bibitem[{{MacNeice} {et~al.}(2004){MacNeice}, {Antiochos}, {Phillips},
  {Spicer}, {DeVore}, \& {Olson}}]{MacNeice2004}
{MacNeice}, P., {Antiochos}, S.~K., {Phillips}, A., {Spicer}, D.~S., {DeVore},
  C.~R., \& {Olson}, K. 2004, \apj, 614, 1028

\bibitem[{{Masson} {et~al.}(2009){Masson}, {Pariat}, {Aulanier}, \&
  {Schrijver}}]{Masson2009}
{Masson}, S., {Pariat}, E., {Aulanier}, G., \& {Schrijver}, C.~J. 2009, \apj,
  700, 559


\bibitem[{Roussev {et~al.}(2012)Roussev, Galsgaard, Downs, Lugaz, Sokolov,
  Moise, \& Lin}]{Roussev2012}
Roussev, I., Galsgaard, K., Downs, C., Lugaz, N., Sokolov, I., Moise, E., \&
  Lin, J. 2012, Nature Physics, 8, 845

\bibitem[{{Savcheva} {et~al.}(2012){Savcheva}, {Pariat}, {van Ballegooijen},
  {Aulanier}, \& {DeLuca}}]{Savcheva2012a}
{Savcheva}, A., {Pariat}, E., {van Ballegooijen}, A., {Aulanier}, G., \&
  {DeLuca}, E. 2012, \apj, 750, 15

\bibitem[{{Schmieder} {et~al.}(2013){Schmieder}, {D{\'e}moulin}, \&
  {Aulanier}}]{Schmieder2013}
{Schmieder}, B., {D{\'e}moulin}, P., \& {Aulanier}, G. 2013, Advances in Space
  Research, 51, 1967

\bibitem[{{Schou} {et~al.}(2012){Schou}, {Scherrer}, {Bush}, {Wachter},
  {Couvidat}, {Rabello-Soares}, {Bogart}, {Hoeksema}, {Liu}, {Duvall}, {Akin},
  {Allard}, {Miles}, {Rairden}, {Shine}, {Tarbell}, {Title}, {Wolfson},
  {Elmore}, {Norton}, \& {Tomczyk}}]{Schou2012}
{Schou}, J., {Scherrer}, P.~H., {Bush}, R.~I., {Wachter}, R., {Couvidat}, S.,
  {Rabello-Soares}, M.~C., {Bogart}, R.~S., {Hoeksema}, J.~T., {Liu}, Y.,
  {Duvall}, T.~L., {Akin}, D.~J., {Allard}, B.~A., {Miles}, J.~W., {Rairden},
  R., {Shine}, R.~A., {Tarbell}, T.~D., {Title}, A.~M., {Wolfson}, C.~J.,
  {Elmore}, D.~F., {Norton}, A.~A., \& {Tomczyk}, S. 2012, \solphys, 275, 229

\bibitem[{{Sun} {et~al.}(2012){Sun}, {Hoeksema}, {Liu}, {Chen}, \&
  {Hayashi}}]{Sun2012b}
{Sun}, X., {Hoeksema}, J.~T., {Liu}, Y., {Chen}, Q., \& {Hayashi}, K. 2012,
  \apj, 757, 149

\bibitem[{{Titov} \& {D{\'e}moulin}(1999)}]{Titov1999}
{Titov}, V.~S. \& {D{\'e}moulin}, P. 1999, \aap, 351, 707

\bibitem[{{T{\"o}r{\"o}k} {et~al.}(2009){T{\"o}r{\"o}k}, {Aulanier},
  {Schmieder}, {Reeves}, \& {Golub}}]{Torok2009}
{T{\"o}r{\"o}k}, T., {Aulanier}, G., {Schmieder}, B., {Reeves}, K.~K., \&
  {Golub}, L. 2009, \apj, 704, 485

\bibitem[{{T{\"o}r{\"o}k} \& {Kliem}(2005)}]{Torok2005}
{T{\"o}r{\"o}k}, T. \& {Kliem}, B. 2005, \apjl, 630, L97

\bibitem[{{T{\"o}r{\"o}k} \& {Kliem}(2007)}]{Torok2007}
---. 2007, Astronomische Nachrichten, 328, 743

\bibitem[{{T{\"o}r{\"o}k} {et~al.}(2011){T{\"o}r{\"o}k}, {Panasenco}, {Titov},
  {Miki{\'c}}, {Reeves}, {Velli}, {Linker}, \& {De Toma}}]{Torok2011}
{T{\"o}r{\"o}k}, T., {Panasenco}, O., {Titov}, V.~S., {Miki{\'c}}, Z.,
  {Reeves}, K.~K., {Velli}, M., {Linker}, J.~A., \& {De Toma}, G. 2011, \apjl,
  739, L63

\bibitem[{{T{\"o}r{\"o}k} {et~al.}(2013){T{\"o}r{\"o}k}, {Temmer}, {Valori},
  {Veronig}, {van Driel-Gesztelyi}, \& {Vr{\v s}nak}}]{Torok2013}
{T{\"o}r{\"o}k}, T., {Temmer}, M., {Valori}, G., {Veronig}, A.~M., {van
  Driel-Gesztelyi}, L., \& {Vr{\v s}nak}, B. 2013, \solphys, 286, 453


\bibitem[{{Wang} \& {Liu}(2012)}]{Wang2012}
{Wang}, H. \& {Liu}, C. 2012, \apj, 760, 101

\bibitem[{{Wiegelmann} \& {Sakurai}(2012)}]{wiegelmann2012solar}
{Wiegelmann}, T. \& {Sakurai}, T. 2012, Living Reviews in Solar Physics, 9, 5

\bibitem[{{Wu} {et~al.}(2006){Wu}, {Wang}, {Liu}, \& {Hoeksema}}]{Wu2006}
{Wu}, S.~T., {Wang}, A.~H., {Liu}, Y., \& {Hoeksema}, J.~T. 2006, \apj, 652,
  800

\bibitem[{{Zuccarello} {et~al.}(2012){Zuccarello}, {Meliani}, \&
  {Poedts}}]{Zuccarello2012}
{Zuccarello}, F.~P., {Meliani}, Z., \& {Poedts}, S. 2012, \apj, 758, 117

\end{thebibliography}

\end{CJK*}
\end{document}